\documentclass[prd,onecolumn,notitlepage,
nofootinbib,aps,tightenlines,
preprintnumbers,amsmath,amssymb,amsfonts,showpacs,superscriptaddress]{revtex4-1}

\usepackage[hyperindex,breaklinks]{hyperref}
\usepackage{graphicx,epstopdf}
\usepackage[utf8]{inputenc}
\usepackage{epsf, array, color}
\usepackage{graphicx}
\usepackage{subfigure}
\usepackage{bbold}
\usepackage{slashed}
\usepackage{amsthm}
\usepackage{amsmath}
\renewcommand{\eqref}[1]{Eq.~(\ref{#1})}

\usepackage{natbib,bm} 
\usepackage{hyperref}
\usepackage{breakurl}
\usepackage{multirow}
\usepackage{bbold}
\usepackage{listings}

\DeclareMathAlphabet{\mathcalligra}{T1}{calligra}{m}{n}

\DeclareMathAlphabet{\mathcalligra}{T1}{calligra}{m}{n}
\newcommand{\beq}{\begin{equation}}
\newcommand{\eeq}{\end{equation}}

\newcommand{\be}{\begin{eqnarray}}
\newcommand{\ee}{\end{eqnarray}}
\newcommand{\bea}{\begin{eqnarray}}
\newcommand{\eea}{\end{eqnarray}}

\newcommand{\MeV}{{~\rm MeV}}
\newcommand{\GeV}{{~\rm GeV}}

\newcommand{\TeV}{{~\rm TeV}}

\newcommand{\GF}{G_{\rm F}}

\newcommand{\nc}{\newcommand}
\nc{\Neff}{N_{\rm eff}}
\nc{\AF}[1]{\textcolor{blue}{[AF: #1]}}
\nc{\nv}[1]{\textcolor{magenta}{#1}}

\nc{\neff}{N_{\rm eff}}
\nc{\Mpl}{M_{\rm pl}}
\nc{\mb}{{~\rm mb}}
\nc{\Li}{$^7${\rm Li }}
\nc{\Tdeut}{T_{\rm deut}}
\DeclareMathOperator\erf{erf}

\begin{document}
\title{BBN for the LHC: constraints on lifetimes of the Higgs portal scalars}

\author{Anthony Fradette}
\affiliation{Department of Physics and Astronomy, University of Victoria, 
Victoria, BC V8P 5C2, Canada}
\affiliation{Perimeter Institute for Theoretical Physics, Waterloo, ON N2J 2W9, 
Canada}
\author{Maxim Pospelov}
\affiliation{Department of Physics and Astronomy, University of Victoria, 
Victoria, BC V8P 5C2, Canada}
\affiliation{Perimeter Institute for Theoretical Physics, Waterloo, ON N2J 2W9, 
Canada}

\date{June, 2017}%

\begin{abstract}
LHC experiments can provide a remarkable sensitivity to exotic metastable massive particles, decaying with significant displacement 
from the interaction point. The best sensitivity is achieved to models where the production and decay occur due to different coupling constants,
and the lifetime of exotic particles determines the probability of decay within a detector. The lifetimes of such particles can be independently limited 
from standard cosmology, in particular the  Big Bang Nucleosynthesis. In this paper, we analyze the constraints on the simplest scalar 
model coupled through the Higgs portal, where the production  occurs via $h\to SS$, and the decay is induced by the small mixing angle of 
the Higgs field $h$ and scalar $S$. We find that throughout the most part of the parameter space, $2 m_\mu < m_S < m_h/2$, the lifetimes 
of exotic particle has to be less than 0.1 seconds, while below  $2m_\mu$ it could grow to about a second. The strong constraints on lifetimes 
are induced by the nucleonic and mesonic decays of scalars that tend to raise the $n/p$ ratio. Strong constraints on lifetimes of the minimal 
singlet extensions of the Higgs potential is a welcome news for the MATHUSLA proposal that seeks to detect displaced decays of exotic particles
produced in the LHC collisions. We also point out how 
more complicated exotic sectors could evade the BBN lifetime constraints. 
 \end{abstract}	
\maketitle

\section{Introduction}

The absence of readily discoverable new physics (NP) at the LHC has presented the physics community with a 
formidable puzzle. While the arguments for NP ``not too far" from the weak scale still loom large, there is a 
distinct desire to explore wider (and wilder) theoretical options away from a simply realized weak-scale supersymmetry, 
or extra space dimensions. One possible strategy to look for new physics is to abandon theoretical 
preconceptions, and start looking for non-standard signatures that the NP could present. 

Large classes of models offer promising avenues 
for a non-standard signal in the production of new exotic particles (possibly of electroweak-scale mass) 
with subsequent decay away from the interaction point (see {\em e.g.} \cite{Graham:2012th,Craig:2015pha,Izaguirre:2015pga,Batell:2016zod}). While both ATLAS and CMS have performed corresponding studies in a variety of contexts 
and for different ranges of displacement \cite{CMS:2014hka,Aad:2015rba,Aad:2015uaa}, it has been recently pointed out that a dedicated and relatively inexpensive detector
\cite{Chou:2016lxi} could extend the physics reach into cases where the decay 
lengths are on the order of $O(100\,{\rm m})$ and beyond.

When both the production and decay of an exotic state $S$ occur through one and the same coupling constant, the chances of detecting such 
NP at the LHC experiments are not great. Indeed, a large displacement implies a very small value for the coupling, which in turn 
leads to very inefficient production rates. Therefore, an ideal case for the collider studies 
would be when the production and decay occur through different 
coupling constants, and $\lambda_{\rm production} \gg \lambda_{\rm decay}$.  For the pair-produced exotics, such a hierarchy can be made 
``natural" as the $\lambda_{\rm decay}\to 0$ limit could lead to an enhanced symmetry. 

If the main signal to search for is an {\em appearance} of abnormal energy deposition or exotic vertex some distance from the interaction 
point, it is then very important to know how small $\lambda_{\rm decay}$ is allowed to be. In more practical terms, one would like to know if 
there is an external to the LHC physics constraint on the lifetimes $\tau_S$ of such exotic particles. An obvious source for such a constraint can be 
early cosmology. The big bang nucleosynthesis (BBN), and its overall agreement with observations 
\cite{Cyburt:2015mya} (apart from the unclear status of $^7$Li) 
can provide a limit on the lifetimes of such particles. In order to derive such limits, one would have to make a 
fairly natural assumption that the Universe was indeed as hot as $T\sim m_S\sim~$electroweak~scale at some point in its history. Subsequent thermal evolution to the BBN temperatures involves self-depletion via $SS\to {\rm SM}$ due to  $\lambda_{\rm production} $, in an expected WIMP-type annihilation process,  and late-time decay of $S\to {\rm SM}$ where depending on lifetimes and the properties of the 
decay products the BBN outcome may get affected. These mechanisms are well-understood in the BBN literature 
(see {\em e.g.} \cite{Jedamzik:2009uy,Pospelov:2010hj} for reviews). We will require that the late decay of $S$ provides a small and acceptable 
perturbation to the standard BBN (SBBN) outcome, which in turn will limit $\tau_S$. 

In this paper, we analyze a fairly minimal model, where a new singlet scalar has predominantly a quadratic coupling to the Higgs boson that regulates both its production at colliders and the intermediate cosmological abundance at $T_{\rm BBN} \ll T \ll m_S$. Given that the 
model is very predictive, it allows to place robust bounds on lifetimes of such particles with a minimum amount of model dependence. 
We find that for most of the analyzed parameter space with $m_S < m_h/2$,  the intermediate abundance of such particles is 
large enough to affect the neutron-proton freeze out ratios at relevant temperatures. This allows us to set fairly robust bounds on
lifetimes of such particles, which come out to be remarkably strong, and shorter than $0.1$ seconds. In what follows we describe the model and 
the cosmological history of $S$ (section II); derive the impact on the BBN (section III); present our results (section IV), and provide related discussion (section V).

\section{The minimal Higgs portal model}

We consider the simplest extension of the SM by a singlet scalar field $S$. 
A new singlet scalar $S$ can have two interaction terms with the Standard Model (SM) at the renormalizable level, in addition to trilinear and quartic self-interactions.
In this scenario, the Lagrangian of the singlet sector (including the SM) generically takes the form
\beq
\mathcal{L}_{H/S} = \mu^2 H^\dagger H - \lambda_H\left(H^\dagger H \right)^2  - V(S) - A S H^\dagger H - \lambda_S S^2 H^\dagger H+ \mbox{kin. terms}.
\eeq
The Higgs expectation value $v= 246~{\rm GeV}$ is assumed to correspond to a global minimum. 
The self-interaction potential $V(S) = \lambda_4 S^4 +\lambda_3 S^3 + \frac{m_{S0}^2}{2} S^2$ can be redefined in such a way that the 
linear term is absent. It is important that the $A,~\lambda_3 \to 0$ and $\langle S \rangle =0$ 
limit would correspond to the case of stable $S$ particles. To simplify the discussion without sacrificing much generality, we 
take $\lambda_{3,4} \to 0$ and assume $Av \ll m_{S0}^2, ~\lambda_S v^2$.

The physical mass of $S$ receives a contribution from the electroweak symmetry breaking,  $m_S =\sqrt{m_{S0}^2 + \lambda_S v^2}$. At linear order in $A$, the mixing angle $\theta$ between physical 
excitations $S$ and $h$ is
\beq
\theta = \frac{Av}{m_h^2-m_S^2}\left(1-\frac{\lambda_Sv^2}{m_S^2}\right).
\eeq
The $\lambda_S$ term arises because the $S$ field develops a small $A$-controlled vacuum expectation value. 
The mixing parameter $\theta$ leads, via the $A$ coupling constant, to the decay of $S$ particles, which can be readily 
derived from
\begin{equation}
{\cal L}_{\rm decay} =  S \times \theta \sum_{\rm SM} O_{h},
\label{Oh}
\end{equation}
where $O_h$ is the set of the standard Higgs interaction terms, with the Higgs field removed: {\em e.g.} $O_h = (m_f/v) \bar ff $
for an elementary SM fermion $f$. 

This Yukawa-type coupling to the SM has been tested in rare meson decays~\cite{OConnell:2006rsp,Pospelov:2007mp,Batell:2009jf,Schmidt-Hoberg:2013hba,Clarke:2013aya} and in proton fixed-target experiments~\cite{Alekhin:2015byh}. The model is mostly ruled out for large mixing angles $\theta \gtrsim 10^{-4}-10^{-2}$ over the \mbox{$m_S \sim$ MeV - 5 GeV} mass range. The proposed experiment SHiP could potential improve current sensitivity down to $\theta \sim 10^{-6}$ for $m_S \sim$~few~GeV~\cite{Alekhin:2015byh}.

In the limit of $\theta \to 0$,  $S$ is stable and could be the dark matter~\cite{Silveira:1985rk,McDonald:1993ex,Burgess:2000yq}. Various limits arise from searches in direct and indirect detection if the particle is stable (see Refs.~\cite{Cline:2013gha,Athron:2017kgt} for recent reviews), but $\lambda_S$ is generically bounded from the constraints on invisible Higgs decay, independently of the direct detection limits. The Standard Model Higgs has a well-predicted decay rate into SM particles of $\Gamma_{SM} = 4.07$ MeV. So far, the properties of 125 GeV resonance are remarkably consistent with the SM Higgs, and therefore there is little doubt that its width is close to $\Gamma_{SM}$. The invisible branching ratio of Higgs decay to $SS$ 
final state is  
\begin{align}
\label{Gh}
\Gamma_{h\to SS} &=\frac{\lambda_S^2 v^2}{8\pi m_h} \sqrt{1-\frac{4m_S^2}{m_h^2}}, \\
Br(h\to SS ) &= \frac{\Gamma_S}{\Gamma_S + \Gamma_{SM}} \simeq 10^{-2} \,\left(\frac{\lambda_S}{0.0015}\right)^2,
\label{Gh1}
\end{align}
where in the last line we assumed $Br(h\to SS) \ll 1$ and $m_S \ll m_h$. The experimental upper bound on the invisible branching ratio of a SM Higgs is 0.19 (at $2\sigma$)~\cite{Belanger:2013xza}, which translates into an upper bound on $\lambda_S$
\beq
\lambda_S \lesssim \frac{0.007}{\left(1-\frac{4m_S^2}{m_h^2}\right)^{1/4}}.
\eeq
If $S$ is to be stable, such small couplings would lead to an excessive abundance of $S$, which invalidates the $Z_2$ symmetric case, 
and forces us to include the decay term. From now on, we will consider $\theta \neq 0$, or in other words the case of unstable $S$ particles. 
 Since our analysis is motivated by the 
LHC physics, we will use $Br(h\to SS )$ as an input parameter, and substitute $\lambda_S$ everywhere employing
(\ref{Gh}) and (\ref{Gh1}).

\subsection{Decay products}

Since $S$ interacts with the SM in the same fashion as the Higgs with an additional $\theta$ mixing factor (\ref{Oh}), its decay properties are similar to those of a light Higgs boson. For the derivations of the actual limits on 
the lifetime of $S$, we need to know its mesonic and nucleonic decay branching ratios. 

The decay channels of a light Higgs have been considered in the early years of the Weinberg-Salam electroweak model~\cite{Ellis:1975ap}, with additional refinements as SM particles, hadronic resonances were being discovered and final-state interactions better understood~\cite{Ellis:1979jy,Raby:1988qf,Truong:1989my}. Hadronic decays in the mass range $2m_\pi < m_S \lesssim 4 \GeV$ are still poorly understood, with models varying by as much as a few orders of magnitude near the di-kaon threshold~\cite{Clarke:2013aya}.

The leptonic decay channels are straightforward, with the decay rate given by
\beq
\Gamma_{S\to l \bar{l}} = \frac{\theta^2}{8\pi} \frac{m_l^2}{v^2} m_S \left(1-\frac{4m_l^2}{m_S^2}\right)^{3/2}. \label{eq:Gamff}
\eeq
If the decaying product is a pair of heavy quarks, there are $\mathcal{O}(1)$ corrections coming from the 1-loop QCD vertex correction~\cite{Cline:2013gha}, which yields the following correction factor~\cite{Drees:1990dq} to the fermionic expressions~(\ref{eq:Gamff})
\beq
f_q = 3\left[1 + \frac{4 \alpha_s}{3\pi} \left(\frac{9}{4}+\frac{3}{2}\ln\frac{m_q^2}{m_S^2}\right)\right]
\eeq
and the factor of three comes from the number of color charges. For better accuracy, we use the higher order perturbative results from the \textsc{HDecay} code~\cite{Djouadi:1997yw} for $m_S > 2.5\GeV$.

Metastable mesons, such as $\pi^\pm$ and $K^\pm,\bar K^0, K^0$ are ``important" decay products, as they can participate in the 
charge-exchange reactions with nucleons and shift the $n-p$ balance, hence affecting the whole nucleosynthetic chain.
In the mass range where the perturbative QCD calculations are no longer valid, we base our baseline calculations on Ref.~\cite{Bezrukov:2009yw}. The scalar-pion interaction can be extracted from the low-energy expansion of the trace of the QCD energy-momentum tensor~(see for {\em e.g.}.~\cite{Voloshin:1985tc,Leutwyler:1989xj}) by integrating out the three heavy quarks and using chiral perturbation theory on the remainder, yielding the effective Lagrangian~\cite{Bezrukov:2009yw}
\beq
\mathcal{L}_{S\pi\pi} = \frac{4}{9}\frac{\theta}{v} S\left(\frac{1}{2} \partial_\mu \pi^0 \partial^\mu \pi^0 + \partial_\mu \pi^+ \partial^\mu \pi^-\right) 
- \frac{5}{3} \frac{\theta m_\pi^2}{v} S \left( \frac{1}{2}\pi^0\pi^0 + \pi^+ \pi^- \right),
\eeq
where we have inserted the SM numerical values for the number of heavy quarks and the first coefficient of the QCD beta function. This leads to  
decay width to charged pions
\beq
\Gamma_{S\to \pi^+ \pi^-} = 2\Gamma_{S\to \pi^0 \pi^0} = \frac{\theta^2}{16\pi}\frac{m_S^3}{v^2} \left(\frac{2}{9} + \frac{11}{9} \frac{m_\pi^2}{m_S^2} \right)^2
\sqrt{1- \frac{4m_\pi^2}{m_S^2}}. \label{eq:LowEPi}
\eeq

\begin{figure}
\centering
 \includegraphics[width= 0.50\columnwidth]{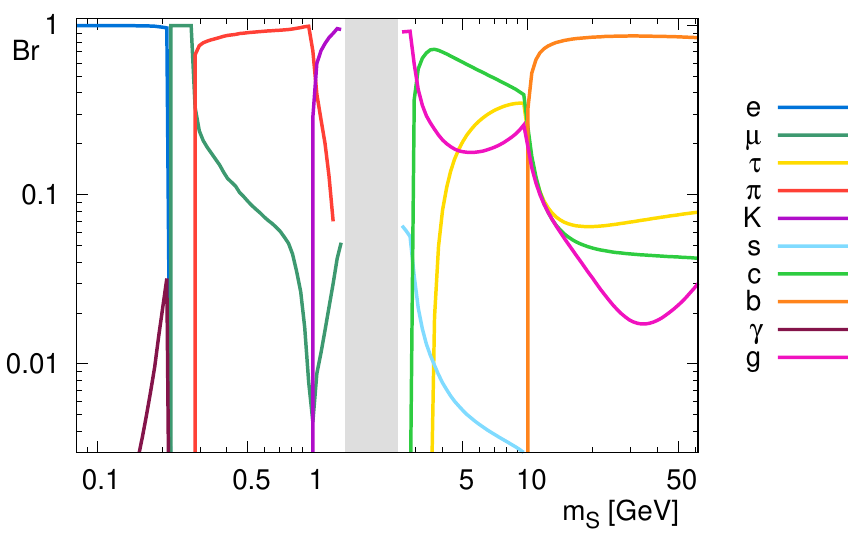} \hspace{0.5cm}
  \includegraphics[width= 0.45\columnwidth]{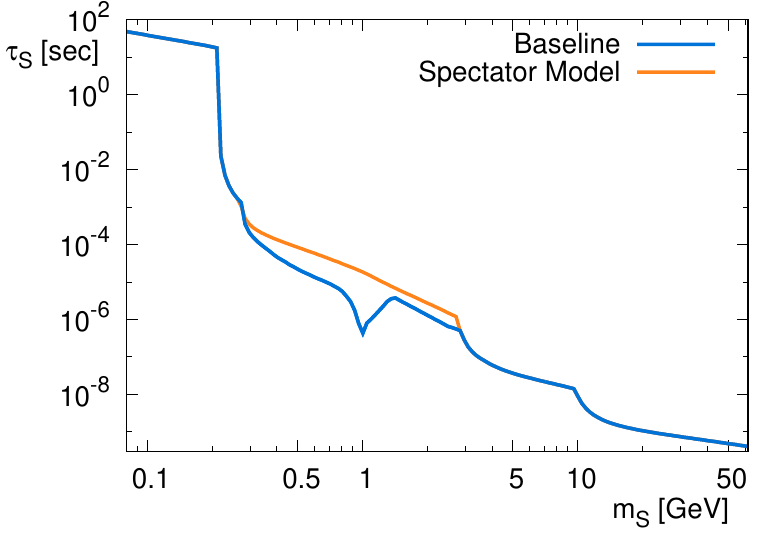}
\caption{\textit{Left}: Branching ratios of the scalar $S$ in our baseline decay model. See text for details. \textit{Right}: Scalar $S$ lifetime of our baseline model and the spectator model for the mixing angle $\theta = 10^{-6}$.} 
\label{fig:S_Br_tau}
\end{figure}

This result is however not applicable far above the pion threshold, as final-state resonances would drastically affect this prediction. Instead, we use the pion and kaon decay width described in Ref.~\cite{Donoghue:1990xh}, where the authors matched the next-to-leading order corrections of the low-energy theorems to the dispersion results from the $\pi \pi$ phase-shift analysis above 600 MeV from the CERN-Munich group~\cite{Hyams:1973zf}. The photon decay channel is added with the prescription detailed in Ref.~\cite{Spira:1997dg}. Finally, there is a gap for $1.4~\GeV < m_S < 2.5~\GeV$ where no analytical treatment is entirely trustworthy, as this includes new resonances strongly coupled to $\eta \eta$ and other potential hadronic cannels. We simply follow Ref.~\cite{Bezrukov:2009yw} and interpolate between the two regimes, under the assumption that there is no order of magnitude deviation in this mass range. The branching ratios and the lifetime for $\theta = 10^{-6}$ are displayed in Fig.~\ref{fig:S_Br_tau}.

As an alternative decay spectrum model, we also display the perturbative spectator approach~\cite{Gunion:1989we,McKeen:2008gd,Alekhin:2015byh}, where the relative decay width above the kaon threshold are given by
\beq
\Gamma_{\mu^+\mu^-} : \Gamma_{KK} : \Gamma_{\eta \eta} = 
m_\mu^2 \beta_{\mu}^3 : 3\frac{9}{13}m_s^2\beta_K^3 : 3\frac{4}{13}m_s^2\beta_\eta^3 , \label{eq:SpectBr}
\eeq
with $\beta_i = \sqrt{1-4m_i^2/m_S^2}\Theta(m_S - 2m_i)$, $\Theta$ being the step-function, and we adopt the running of $s$ quark mass following Ref.~\cite{Spira:1997dg}. The pion contribution is kept as in equation~(\ref{eq:LowEPi}) and then we use the \textsc{HDecay} output at the $c$-quark threshold and above to match our baseline model. 

For $m_S$ of several GeV and heavier, decays with final state nucleon-antinucleon pairs are possible. Even though the 
branching to such states are generally lower than 10\%, the effect on BBN can be quite significant, and therefore these are by 
far the most important channels for $\tau_S \gtrsim 1\;\sec$. On top of direct and for the most part subdominant contributions 
from $S\to \bar nn,...$, we need to take into account the (anti-)nucleon states that emerge from the hadronization of the 
quark decay products and heavy $B$-meson fragmentations. 

\subsection{Cosmological metastable abundance}

\begin{figure}
\centering
 \includegraphics[width= 0.45\columnwidth]{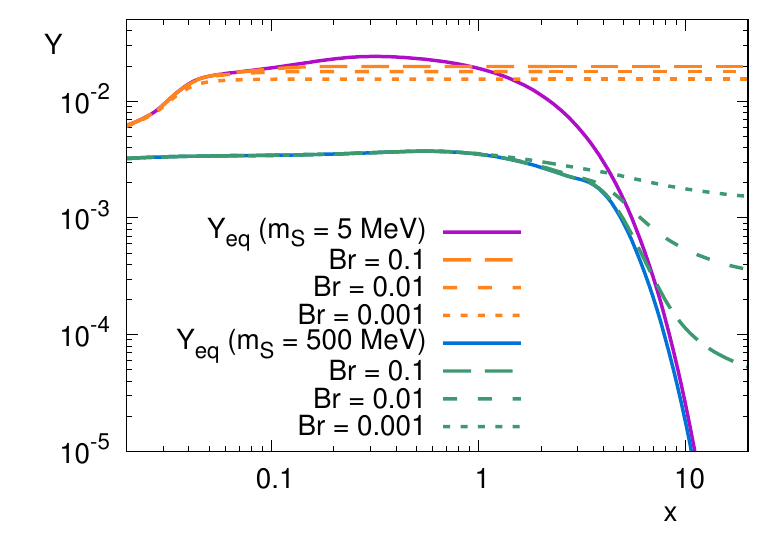} \hspace{1cm}
  \includegraphics[width= 0.45\columnwidth]{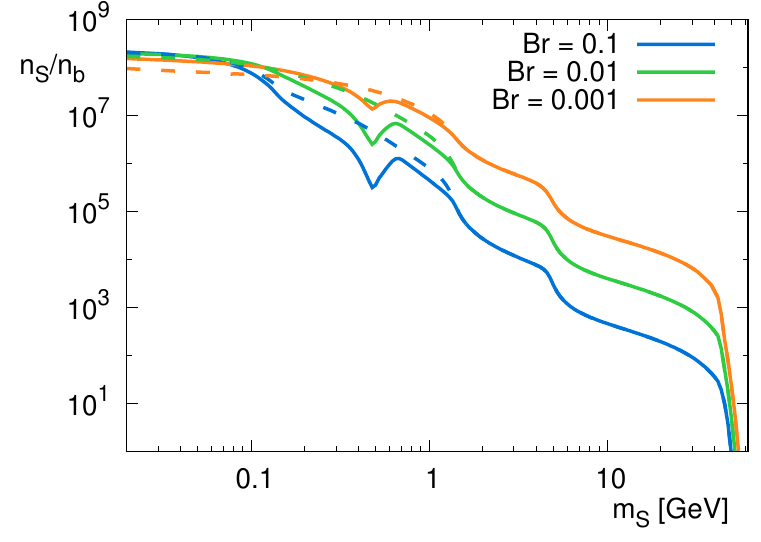}
 \caption{\textit{Left}: Temperature evolution ($x=m/T$) of the $Y_S$ intermediate abundance for $m_S = 5\MeV$ and 500 MeV for the three benchmark higgs branching ratios. \textit{Right}: Metastable abundance of $S$ prior to its decay normalized over the baryon density. Values shown for $Br(h\to SS) = 10^{-1}$, $10^{-2}$ and $10^{-3}$. The dashed lines correspond to the perturbative spectator model.} 
\label{fig:YS_abundance}
\end{figure}

After the temperature drops below $m_S$, the interaction of $SS$ pairs with the SM shifts towards the annihilation, 
resulting in an intermediate (metastable) population of $S$ bosons.
In the mass range that we consider, the $S$ annihilation is dominated by the $s$-channel reactions $SS \to h^* \to XX$, where on the 
receiving end are the pairs of the SM states $XX$ created by a Higgs-mediation process. The annihilation cross section $\sigma v$ generically takes the form
\beq
\sigma v(s) = \frac{8 \lambda_S^2 v^2}{(s-m_h^2)^2 + m_h^2 \Gamma_{{\rm SM}+S}^2}\frac{\Gamma_{{\rm SM}}^{m_h\to\sqrt{s}}}{\sqrt{s}}, \qquad \qquad
\langle \sigma v \rangle = \frac{\int_{4m_S^2}^\infty ds\;  \sigma v (s)\;s \sqrt{s-4m_S^2}K_1\left(\frac{\sqrt{s}}{T}\right)}{16 T m_S^4 K_2^2\left(\frac{m_S}{T}\right)}.
\label{eq:sigmav}
\eeq
This formula recast the rate in terms of a Higgs width $\Gamma_{{\rm SM}}^{m_h\to\sqrt{s}}$ with a fictitious mass of $\sqrt{s}$.   This form encompasses both perturbative and non-perturbative channels in the $h^*$ decay rate (with the substitution $m_h^*\to \sqrt{s}$), which we have described above. In the standard WIMP freeze out paradigm, a DM particle freezes out at $T_{\rm f.o.} \sim m_{DM}/20$, $\langle \sigma v \rangle$ is simply the nonrelativistic limit $\sigma v (\sqrt{s} =2 m_{DM})$ and the relic density can be conveniently approximated as $\Omega_{DM} h^2 \sim 0.11\times 1{\rm pb}/\langle \sigma v \rangle$. This result emerges as a solution to the Boltzmann equation\footnote{We use the standard variable definitions, where $Y = n_S/ s$ is the $S$ abundance normalized on the entropy density $s$, $x = m/T$ is the dimensionless inverse temperature, $H$ is the Hubble rate, $h_{\rm eff}$ is number of entropic relativistic degrees of freedom and $Y_{eq}$ is the normalized thermal equilibrium $S$ number density.}~\cite{Steigman:2012nb}
\beq
\frac{dY}{dx} = \frac{s\langle \sigma v \rangle}{Hx} \left [ 1 + \frac{1}{3} \frac{d(\ln h_{\rm eff})}{d (\ln T)} \right ] \left( Y_{\rm eq}^2 - Y^2\right), \label{eq:BoltzmannRelic}
\eeq
when the freeze out occurs in the exponentially falling region of the equilibrium density $Y_{\rm eq}(T)$. For a much smaller annihilation cross 
section, $\langle \sigma v \rangle \ll 1\;{\rm pb}$, $Y$ departs from the equilibrium value earlier, possibly near the relativistic plateau $Y_{\rm eq} = n_{\rm eq}/ s \to 45 \zeta (3) /2 \pi^4 h_{\rm eff} (T)$ for $x \ll 1$. Since the nonrelativistic annihilation cross section in the minimal Higgs portal model ranges from $10^{-3}$ to $10^{-14}$ pb for $m_S \sim 1\MeV - 60 $ GeV and \mbox{$Br(h\to SS) \sim 0.1 - 0.001$}, we numerically integrate equation~(\ref{eq:BoltzmannRelic}) to determine the metastable $S$ abundance. The results are shown in Fig.~\ref{fig:YS_abundance}, normalized to the baryon number density for a more intuitive interpretation of its impact on BBN in the following section.

For $m_S \simeq m_h/2$, the  $\sigma v$ cross section evaluated at $s=4m_S^2$ is a poor approximation, as it fails to capture the strong energy dependence of the cross section near the resonance at $\sqrt{s} = m_h/2$ \cite{Griest:1990kh}. The sharp drop in the abundance above $m_S \sim 45\GeV$ is due to the resonant contribution to the thermally averaged cross section, leading to a delayed freeze out and drastic decrease in metastable $S$ abundance. Our numerical results agree with the semi-analytic treatment of Ref.~\cite{Cline:2013gha}. For very light $m_S$, one can see that the freeze out abundances are large, and the relative spread between different input values of $Br(h\to SS)$ gets smaller, as the annihilation cross section becomes very small and the freeze out happens in the semi-relativistic regime $x_{\rm f.o.} \sim \mathcal{O}(1)$ and asymptote to the $Y_{\rm eq}$ relativistic plateau for small $m_S$. The only difference at the lightest masses is from $Y_{\rm eq}^{rel} \propto 1/h_{\rm eff}(T)$. Since $h_{\rm eff}$ is a monotonic function of temperature, weaker annihilation cross sections freeze out earlier, at a higher temperature,  thus yielding smaller abundances (as seen in the $m_S = 5\MeV$ curves in Fig.~\ref{fig:YS_abundance}). This is in contrast with the standard freeze out in the non-relativistic regime, with final abundances inversely proportional to the cross section.  
We note in passing that the strong-interaction-related uncertainty ``propagates" outside the  $m_S \sim 2m_\pi - 2 m_c$ window. For example, because of the relativistic freeze out, for $m_S$ smaller $ 2 m_\pi$ the hadronic channels may turn out to be important.

\section{Big Bang Nucleosynthesis}

The formation of light nuclei is one of the earliest probes of NP in cosmology along with far less certain constraints imposed by the inflationary framework. BBN is well-understood within SM physics, and its outcome agrees with observational data for $^4$He and D. $^7$Li has an outstanding factor of $\sim 2-3$ discrepancy between theory and observations~\cite{Cyburt:2015mya}, with the caveat that the observed abundances 
may have been affected by stellar evolution. Nevertheless, the overall success over a wide range of abundances can be used to constrain various types of NP~\cite{Pospelov:2010hj}.

The initial BBN stage is the neutron-proton ratio $n/p$ freeze out. Maintained in equilibrium by electroweak interactions at high temperatures, the neutron abundance follows $n/p \sim e^{-Q/T}$, where $Q=m_n - m_p -m_e \simeq 1.293\MeV$, until the epoch when the weak processes decouple around temperatures of 0.7 MeV. The outcome, $n/p \simeq 1/6$, is quasi-stable, decreasing to $n/p \simeq 1/7$ at the end 
of the ``deuterium bottleneck". The latter terminology is used to indicate a much delayed onset of nuclear reactions controlled by a relatively shallow $n-p$ binding energy. 
Once the Universe runs out of photons that can efficiently dissociate deuterium, the bulk of the nucleosynthetic 
reactions occurs at $t_{\rm deut} \sim 200 $ seconds. $^4$He has a large binding energy per nucleon, and the reactions leading to it are less Coulomb-suppressed than for heavier elements. Consequently, 
most neutrons end up in the final $^4$He abundance (expressed in mass fraction from the total baryon mass) $Y_p \simeq 2 (n/p)/\left( 1 + n/p\right) \simeq 0.25$. 

Traces of neutrons and incomplete nuclear burning of $A=2,~3$ nuclei light nuclei
result in the left-over abundances of $^3$He and D. 
Beyond the $^4$He atomic number, the deepest bound nucleus is $^{12}$C, but its formation is completely 
suppressed since it would need to be produced by a triple $^4$He collision. The $2\to2$ reactions $p+{}^4$He and $^4$He + $^4$He are also ineffective at producing heavier nuclei as the $A = 5$ and $A = 8$ elements are all unstable. The only remaining possibilities are $^4{\rm He}+{}^3{\rm He} \to {}^7{\rm Be} + \gamma$ followed by a $\beta$ decay to yield $^7{\rm Li/H} \sim \mathcal{O}(10^{-10})$ and $^6$Li formed at the $^6{\rm Li/H} \sim \mathcal{O}(10^{-14})$ via $^4$He-D fusion.
For the problem at hand - the determination of the upper limit on the $S$ lifetime - few of these details matter. This is because of relatively large 
metastable abundances affecting the earliest stages of nucleosynthesis, primarily via the $n/p$ ratio.

\subsection{Neutron Enrichment}

Ample abundances of $S$ particles ($n_S \sim 10^{2}-10^{9} \times n_b$) flood the Universe with final state mesons and nucleons
that in turn could spoil the final light nuclei abundances. For example, at temperatures $T\sim 0.5 $ MeV, the protons are 
$\sim6$ times more abundant than neutrons, but this ratio can be easily changed due to meson-induced charge exchange reactions. 
At these temperatures, the probability of $p\to n$ conversion from charged pions is
\beq
P_{n\to p} \simeq n_p \langle \sigma v \rangle_{pn} c\tau_{\pi^+} \simeq 2\times \frac{10^{21}}{\rm cm^3}\times 1.5 \mb \times 2.6 \times 10^{-8}{~\rm sec}\times c \simeq 2.5\times 10^{-3}.
\eeq
It is then clear that injection of $O(10^3)$ mesons per nucleon at these temperatures can drastically increase the $n/p$ freeze out abundance. 
Similarly, direct baryonic injection of $n\bar{n}$ and $p\bar{p}$ will have a similar effect on the $n/p$ ratio. On the other hand, if $S$ decays happen before the $n/p$ freeze out, the additional $p\to n$ conversions would not be as efficient,  being washed out by the on-going weak interaction conversions. 

The limit of the exclusion region in the $Y_S/\tau_S$ parameter space ($Y_S \equiv n_S/n_b$ from now on) is determined by solving the Boltzmann equation with the injection of charge exchange inducing particles. Given that the abundances of $S$ particles are large, the main constraints can be derived from the $n/p$ freeze out ratio.
To that effect, we would not need a complete BBN framework, but only a subset of the whole code that deals with $n\leftrightarrow p$ conversions.     We follow the semi-analytic treatment by Mukhanov \cite{Mukhanov:2003xs}, that approximates
$n\leftrightarrow p$ weak conversion rates by a few integrals over thermal distributions, and assumes a ``step-like" disappearance of charged leptons below $T=m_e$,
\begin{align}
\Gamma_{n \nu_e \to p e^- } &= \frac{1+3g_a^2}{2\pi^3} G_F^2 Q^5 J(1;\infty),  &\Gamma_{ p e^- \to n \nu_e } &=e^{-Q/T} \Gamma_{n \nu_e \to p e^- }, \qquad \\
\Gamma_{n e^+ \to p \bar{\nu}_e } &= \frac{1+3g_a^2}{2\pi^3} G_F^2 Q^5 J(-\infty;-\frac{m_e}{Q}), &\Gamma_{p \bar{\nu_e} \to n e^+ } &=e^{-Q/T} \Gamma_{n e^+ \to p \bar{\nu}_e }, \\
J(a,b) &\equiv \int_a^b \sqrt{1-\frac{(m_e/Q)^2}{q^2}}\frac{q^2(q-1)^2\;dq}{(1+e^{\frac{Q}{T_\nu}(q-1)})(1+e^{-\frac{Q}{T}q})} ,
\end{align}
where $g_a \simeq 1.27$ is the standard nucleon axial-vector coupling, $Q = m_n - m_p - m_e \simeq 1.293\MeV$, and $G_F$ is the Fermi constant.  The reverse reaction rates are found by detailed balance. We evaluate $J$ numerically and solve for the electron-neutrino temperature $T_\nu$ by entropy conservation, assuming a $\nu_e$ decoupling temperature of 2 MeV, which reproduces the correct entropy degrees of freedom at lower temperature~\cite{Srednicki:1988ce}. It is then straightforward to solve numerically the differential equation for $X_n = n_n/n_b$,
\beq
\frac{d X_n}{dT} = \frac{\Gamma_{n \nu_e \to p e^- }+\Gamma_{n e^+ \to p \bar{\nu}_e } }{T H(T)} \left(X_n -(1-X_n)e^{-Q/T} \right) +
\frac{\Gamma_n X_n}{T H(T)}, \label{eq:Boltzmann}
\eeq 
where the last term represents the neutron decay with $\Gamma_n^{-1} = 880~{\rm sec}$.  
This equation is approximately valid until the rapid switch-on of the nuclear reaction rates at the end of the deuterium bottleneck. 
Within this approximation, one can determine the final temperature where the equation is valid by starting with $X_n = 1/2$ at early times, and solving for the deuterium bottleneck temperature by imposing $Y_p = 2 X_n (\Tdeut)= 0.25$. This results in $\Tdeut \simeq 0.068 \MeV$ or $t_{\rm deut} \simeq 276$ sec. We take this approximation as our baseline SBBN model, which is then modified by the inclusion 
of extra sources and sinks for $n,~p$, and new $n\leftrightarrow p$ reactions. 
To constrain the parameter space of a species decaying into charged mesons or baryons, we proceed by solving the Boltzmann equation that includes new interactions. We will require that $Y_p$ does not deviate from SBBN by more than 4\%, 
\beq
\Delta Y_p \equiv |Y_p -Y_p^{\rm SBBN}| < 0.01,
\eeq 
which is a rather generous allowance for the errors, considering the tight observational constraints on primordial helium abundance 
\cite{Cyburt:2015mya}. Consequently, it will result in conservative limits of $\tau_S$. 

\subsubsection{Meson-Mediated mechanism}

Only long-lived mesons have an opportunity to interact with the baryon bath and induce proton-neutron conversions. As such, only $\pi^\pm$, $K^\pm$ and $K_L$ have lifetimes in excess of $\tau \sim 10^{-8}$ seconds, and can induce $p\leftrightarrow n$ via strong interactions. For temperatures relevant for the $n/p$ freeze out, the density of charged leptons is very high, and mesons are efficiently stopped 
by the primordial plasma. We assume that they are efficiently thermalized, and take the relevant pion-induced reactions at threshold~\cite{Reno:1987qw,Pospelov:2010cw} ($c=1$),
\begin{align}
\pi^- + p &\to n + \gamma : &(\sigma v)^{\pi^-}_{pn(\gamma)} &\simeq 0.57 \mb, &Q &= 138.3\MeV,\\
\pi^- + p &\to n + \pi^0 : &(\sigma v)^{\pi^-}_{pn(\pi^0)} &\simeq 0.88 \mb, &Q &= 3.3\MeV,\\
\pi^+ + n &\to p + \gamma : &(\sigma v)^{\pi^+}_{np(\gamma)} &\simeq 0.44 \mb, &Q &= 140.9\MeV,\\
\pi^+ + n &\to p + \pi^0 : &(\sigma v)^{\pi^+}_{np(\pi^0)} &\simeq 1.26 \mb, &Q &= 5.9\MeV.
\end{align}
The reverse reactions are irrelevant due to the short lifetime of  $\pi^0$'s and the need for non-thermal $\gamma$'s of $\sim 140$ MeV energy. The $\pi^-$ reactions are to be added to the r.h.s of  Boltzmann equation~(\ref{eq:Boltzmann}) via the additional term
\beq
\left.\frac{d X_n}{dT}\right|_{\pi-} = \frac{-1}{TH(T)}n_{\pi^-}^{\rm inj} \left( \langle\sigma v \rangle_{pn(\pi^0)}^{\pi^-}+ \langle\sigma v \rangle_{pn(\gamma)}^{\pi^-}\right)(1-X_n) ,
\label{eq:dXdTpi}
\eeq
and similarly for the $\pi^+$ reactions. The ambient population of injected pions from a $S$ decay with \mbox{$Br(S\to \pi^+\pi^-)=\xi_{\pi^\pm}$} is $n_{\pi^\pm}^{\rm inj} \simeq \xi_{\pi^\pm}\Gamma_S \tau_{\pi^\pm} Y_S n_b(T) e^{-t\Gamma_S}$, $t \simeq 2.42\sec\; ({\rm MeV}/T)^2/\sqrt{g_\star}$
 and the thermal cross section are taken at their threshold value $\langle \sigma v \rangle_{np}^{\pi^+}=(\sigma v)_{np}^{\pi^+}$.
 Reactions with pairs of charged particles in the initial states, such as $\pi^-p$, will be somewhat enhanced due to the 
 Coulomb attraction, which provides a small but non-negligible correction. We account for it following Ref.~\cite{Pospelov:2010cw}.

\begin{figure}
\centering
 \includegraphics[width= 0.41\columnwidth]{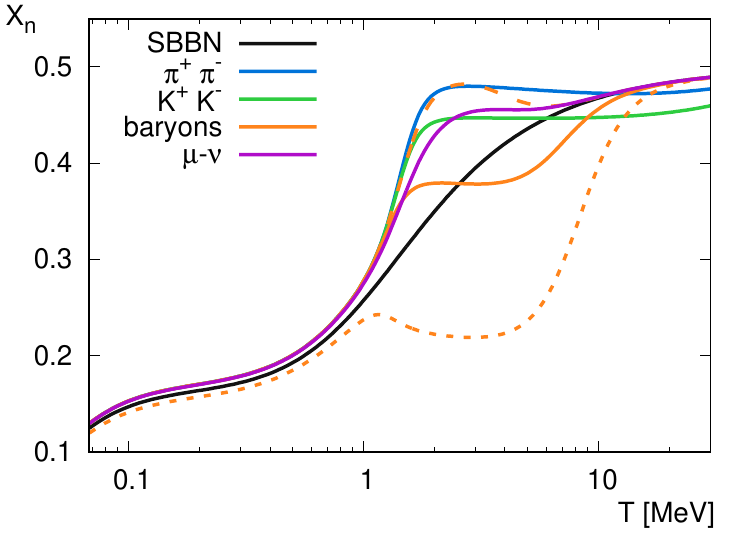} \hspace{1cm}
  \includegraphics[width= 0.45\columnwidth]{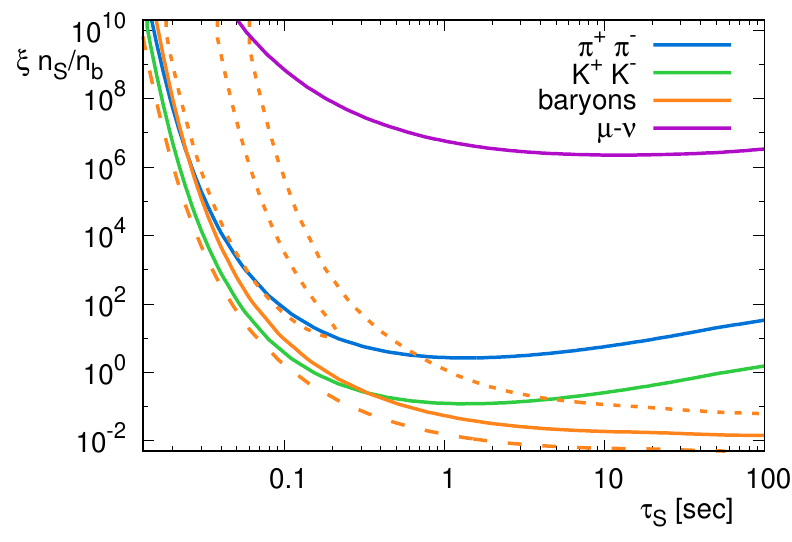}
\caption{\textit{Left}: $X_n$ evolution for the SBBN and the injection of pions, kaons, baryons and muons (neutrinos) as described in the text for lifetimes of $0.05$ seconds with the initial $Y_S$ abundance tuned to yield $\Delta Y_p=0.01$ (maximum allowed shift of $Y_p$). The baryonic injection is taken at $\kappa = 0.5$ (full line), the lines for $\kappa = 1$ (dashed) and $\kappa = 0.2$ (dotted) are also displayed. \textit{Right}: Limit of injected pairs for each channel as a function of the $S$ lifetime. The upper-right dotted line for $\kappa = 0.2$ is at $Y_p = 0.26$, the upper-left dotted island yields $Y_p = 0.24$. } 
\label{fig:Xn_Gamtau}
\end{figure} 

The implementation of the charged kaons reactions is similar to the pion case, but the dominant reactions are rather different. The direct charge exchange between neutral and charged kaons is 
\beq
\bar{K}^0 + n \to K^- + p: \qquad (\sigma v)^{\bar{K}^0}_{pn(K^-)} \simeq 10 \mb, \qquad Q = 5.3\MeV, \label{eq:Kmp}
\eeq
with similar cross section for a charge-conjugated reaction, $K^0p\to K^+n$. For neutral kaons, the effects induced by 
$K_L$ are the most important, and we use $\sigma (K_L n \to K^- p ) \simeq \frac{1}{2}\sigma (\bar{K}^0n \to K^-p)$ and~(\ref{eq:Kmp}) to find $(\sigma v)^{\bar{K}^0}_{pn(K^-)} \simeq 4.5 \mb$.
Additionally, efficient reactions can also proceed via $s$-quark being incorporated inside a hyperon that subsequently decays into $p/n$ + $X$. 
The inclusive threshold cross section found by weighting each hyperon with their branching ratios to $p/n$ are~\cite{Pospelov:2010cw}
\begin{align}
K^- + p &\to n + X : &(\sigma v)^{K-}_{pn} &\simeq 32\mb, \\
K^- + n &\to p + X : &(\sigma v)^{K-}_{np} &\simeq 13\mb, \\
K_L + p &\to n + X : &(\sigma v)^{K-}_{pn} &\simeq 6.5\mb, \\
K_L + n &\to p + X : &(\sigma v)^{K-}_{np} &\simeq 16\mb.
\end{align}
Notice the absence of corresponding hyperon reactions initiated by $K^+$ due to the presence of anti-$s$ quark. 

Representative examples of  $X_n(T)$ evolutions and the sensitivity to $\xi_{\pi^\pm}Y_S/\tau_S$ parameter space are shown in Fig.~\ref{fig:Xn_Gamtau}. Left panel displays significant modifications to the evolution of neutron abundance at $\tau_S =$0.05~seconds with adjustable initial abundance, yielding $\Delta Y_p = 0.01$. The departure 
from $X_n = 0.5$ at high temperatures is clearly visible. (In fact, for short $\tau_S$, the kaon injection channel at early times leads to a shift of the equilibrium value of $X_n$ 
to $(\sigma v)^{K^-+K_L}_{pn}/((\sigma v)^{K^-+K_L}_{pn}+(\sigma v)^{K^-+K_L}_{np}) \simeq 0.45$.) 
As the temperature lowers, the Coulomb-enhanced reaction becomes stronger. For meson injection, these reactions enhance the $p\to n$ conversion, keeping $X_n$ away from the SBBN value. 
Right panel gives a boundary of the exclusion regions for different 
injection modes.
In addition to the already described channels, charged kaons also give rise to a population of secondary charged pions that should also be included in the analysis of $p\leftrightarrow n$ transitions. Since the constraints are already stronger than for 
the charged pion case, we neglect this effect, which leads to more conservative bounds.

\subsubsection{Direct baryonic injection mechanism}

If $S$ is heavy enough, the end-products after hadronization of the primary decay products ({\em e.g.} $b$ or $c$ quarks) 
may contain baryons. Since $S$ has no baryon number,  one should expect an equal number of baryons and anti-baryons in the final states. Therefore, one should expect the injection of $n\bar{n}$, $p\bar{p}$, $\bar n p$ and $p \bar n$ pairs, as well as (in principle) baryonic states with higher multiplicities. 
The hadronization process and decay of heavy quarks 
produce much more light mesons than baryons, and a complete analysis must include a Monte Carlo study of the hadronization process (see Ref.~\cite{Kohri:2001jx} for benchmarks of heavy unstable particles decaying into 2 hadronic + 1 leptonic jets in the early BBN epoch). 
Assuming that the heavy quarks inside baryons decay due to the ``main" weak decay sequence, $b\to c\to s \to u$, one should also expect a 
somewhat large number of the final states with a proton or anti-proton over neutron or anti-neutron. 
We will tune the branching models of $S$ to available particle data on proton production, and 
take $N_{n} = \kappa N_{p}$ and $N_{\bar n} = \kappa N_{\bar p}$. Furthermore, due to a more frequent 
appearance of up-quark at the end of the decay chain, we would take $\kappa \simeq  0.5$  on average.

As in the case of mesons, the thermalization of baryonic decay products is quick (see {\em e.g.}~\cite{Kawasaki:2004qu}). As a baryonic pair is created in the decay, the baryon is added to the existing population of $n$ or $p$. The anti-baryon will, however, annihilate with either 
$p$ or $n$ and dissipate into lighter mesons. If it annihilates with its own antiparticle, there is no net change in $n/p$, but an annihilation with the other species induces a net $n-p$ change. The probability $P_{k\bar{l}}^{i\to j}$ of a net charge exchange $i\to j$ from a $k\bar{l}$ injection is simply given by the weighted annihilation rates
\beq
P_{p\bar{p}}^{n\to p} = \frac{X_n \langle \sigma v \rangle_{n\bar{p}}}{X_n \langle \sigma v \rangle_{n\bar{p}}+ (1-X_n)\langle \sigma v \rangle_{p\bar{p}}}, 
\qquad 
P_{p\bar{n}}^{n\to p} = \frac{(1-X_n) \langle \sigma v \rangle_{p\bar{n}}}{X_n \langle \sigma v \rangle_{n\bar{n}}+ (1-X_n)\langle \sigma v \rangle_{p\bar{n}}}
\eeq
and similarly for the $n\bar{p}$ and $n\bar{n}$ injections. The baryonic annihilation rates are given by ~\cite{Reno:1987qw}
\beq
\langle \sigma v \rangle_{n\bar{n}} = \langle \sigma v \rangle_{p\bar{p}}  / C = 37\mb, \qquad
\langle \sigma v \rangle_{n\bar{p}} = \langle \sigma v \rangle_{p\bar{n}} = 28\mb, 
\eeq
where the $p\bar{p}$ has the low-$v$ Coulomb correction $C(v)$. The implementation of these processes in the Boltzmann equation then
require additional terms
\beq
\left.\frac{d X_n}{dT}\right|_{pn} = \frac{-\xi_p \Gamma_S e^{-t\Gamma_S}}{TH(T)}\left( -P_{p\bar{p}}^{n\to p}  - \kappa P_{p\bar{n}}^{n\to p} + \kappa P_{p\bar{n}}^{n\to p} + \kappa^2 P_{n\bar{n}}^{n\to p}  \right).
\eeq
As before, the outcome is displayed in Fig.~\ref{fig:Xn_Gamtau}. Again, for short $S$ lifetimes and large $Y_S$, the large numbers of injected particles completely dictates the early $X_n$ value. The constraint on $Y_S$ goes up more sharply in the short $S$ lifetime limit. There is a significant dependence on $\kappa$ for $\tau_S \gtrsim 0.1$ sec, which is washed out by the SM electroweak interactions at earlier times. If we take the extreme limit $\kappa \to 0$, no neutrons are injected and the $p\bar{p}$ pair can only further decrease the $n/p$ ratio, thus constrained by the lower $Y_p$ limit 0.24. On the other hand, a symmetric injection $\kappa =1 $ enhances the $n/p$ ratio as the anti-baryon mostly annihilates on protons, more abundant than neutrons by a factor of $\sim 6-7$ after the standard $n/p$ freeze out. For $\kappa \gtrsim 0$, the final $Y_p$ can either be increased or decreased, depending if the $S$ particles decay away before or after the displaced $X_n$ equilibrium crosses the SBBN $n/p$ freeze out curve. As shown for $\kappa = 0.2$ in Fig.~\ref{fig:Xn_Gamtau}, there is a $Y_p = 0.24$ exclusion island at low lifetimes and larger lifetimes are constrained by $Y_p = 0.26$. We use $\kappa = 0.5$ as our baryonic injection constraint benchmark.

\subsubsection{Muon-Mediated mechanism}

Muon injection physics differs from the previous scenarios of meson and baryon injection. The direct charge-exchange is through the weak force, as opposed to the strong force in the other cases, and is completely negligible over the lifetime of the muon. Instead, the reactions can happen via the energetic neutrinos emitted by the muon decays. The case for muon injection after $t \sim 100\;\sec$ has been covered in Ref.~\cite{Pospelov:2010cw}, to which we refer the reader for details. Assuming stopped muons, the authors solved for the injected neutrino energy spectrum, including redshifting and averaged over flavour oscillations, to be integrated in the $n-p$ conversion rate. At earlier times, we know background neutrinos are coupled to $e^\pm$ down to $T\simeq2\MeV$, and energetic injected neutrinos must accordingly deplete their energy efficiently as well. Summing over the possible interactions with the background neutrinos and $e^\pm$~\cite{Dolgov:1997mb}, the collision rate of an injected electron-neutrino with the bath is given by 
\begin{align}
\nonumber
\Gamma^{\nu_e}_{\rm coll} \left(E_\nu , T \right) &= \frac{7\pi}{135}  \GF^2 E_{\nu} \left[ \left(5 + g_{\rm L}^2 + g_{\rm R}^2\right) T_\nu^4 + 4 \left(g_{\rm L}^2 + g_{\rm R}^2 \right) \eta(T) \;T_\gamma^4 \right], \label{eq:Collnue}\\
&\simeq \left( \frac{E_\nu}{32\MeV}\right) \left[ \frac{5.7}{\sec} \left(\frac{T_\nu}{1\MeV}\right)^4 +\frac{1.3}{\sec} \eta(T_\gamma)\left(\frac{T_\gamma}{1\MeV}\right)^4
\right],
\end{align}
$g_{\rm L} = 1/2 + \sin^2\theta_{\rm w}$, $g_{\rm R} = \sin^2\theta_{\rm w}$, while $\eta(T) = 1$ for $T\gtrsim m_e$ and exponentially falls to 0 at lower temperatures. We follow the implementation of Ref.~\cite{Pospelov:2010cw} and correct for the removal of energetic neutrinos by adding an effective collision lifetime in the neutrino energy distribution (normalized on $n_b$) 
\beq
f_e\left(T,E_\nu\right) = \Gamma_S Y_S \int_T^\infty \frac{dT_1\; e^{-t_1\Gamma_S}}{H(T_1)T_1} F_e \left(E_\nu , \frac{E_0 T}{T_1}\right) e^{-\int_T^{T_1} dT_2\; \frac{\Gamma_{\rm coll}(E_\nu \frac{T_2}{T}, T_2)}{H(T_2)T_2}},
\eeq
where $F_e$ is the distribution at injection time $T_1$, averaged over flavour oscillations. The charge-exchange rate to be inserted in the Boltzmann equation~(\ref{eq:Boltzmann}) is 
\beq
\Gamma_{pn}^\nu = n_b (T) \int_0^{E_0} \sigma_{pn}^{\bar{\nu}} f_e (T,E_\nu)\; dE_\nu
\eeq
and similarly for the reverse $np$ direction. The resulting constraints are shown in Fig.~\ref{fig:Xn_Gamtau}. Our results lean on the conservative side on a few assumptions. For simplicity, we assumed one collision for the neutrino thermalization, instead of following energy degradation over a shower of multiple interactions. Moreover, we took the collision time of the electron-neutrino, even though there are muon-neutrino states in the oscillations. Since $\Gamma_{\rm coll}^{\nu_e} > \Gamma_{\rm coll}^{\nu_\mu}$, we over-estimate the actual collision time and the overall conversion rate should be slightly larger.

\subsubsection{Meson injection from residual annihilations}

\begin{figure}
\centering
 \includegraphics[width= 0.48\columnwidth]{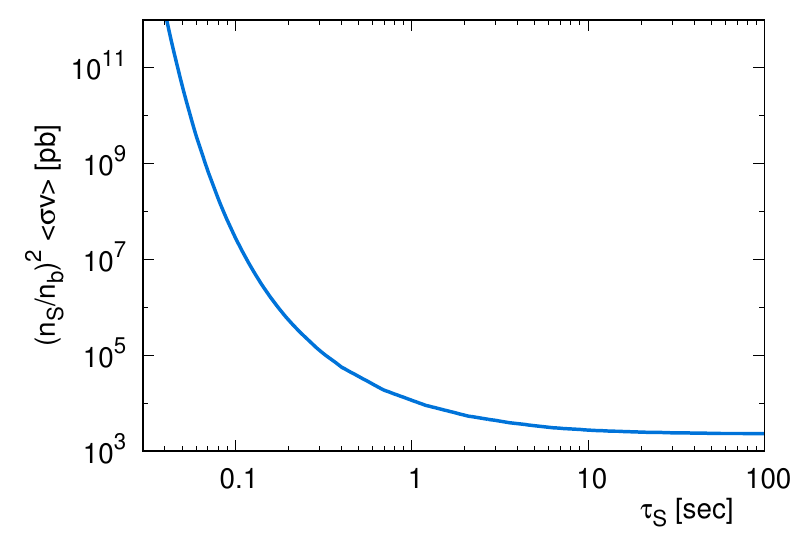} 
\caption{Constraints on $Y^2_S \langle \sigma v\rangle_{\pi^+\pi^-}$ from $SS$ annihilations into charged pions from the BBN $^4$He abundance at $Y_p = 0.26$.} 
\label{fig:Y2sigann}
\end{figure}

In addition to its decay products, $S$ can also inject particles in the cosmic medium via $SS$ annihilations to charged pions. The injected pions interact with the cosmic medium in the same way as from $S$ decays described above. The Boltzmann equation takes the addition term~(\ref{eq:dXdTpi}), with the injected pion density now given by
\beq
n_{\pi^\pm}^{\rm ann} = \tau_{\pi^\pm}  n_S^2(T) \left\langle \sigma v \right \rangle_{\pi^+\pi^-} = \tau_{\pi^\pm}  Y_S^2 n_b^2(T) e^{-2t\Gamma_S}  \left\langle \sigma v \right \rangle_{\pi^+\pi^-},
\eeq
where $\langle \sigma v \rangle_{\pi^+\pi^-}$ is the non-relativistic annihilation cross section $\sigma v (2m_S)$~as per Eq.~(\ref{eq:sigmav}), rescaled by the pionic branching ratio at $\sqrt{s}= 2 m_S$. The $n_{\pi^\pm}^{\rm ann} \propto n_S^2 \propto T^6$ dependence imply a much stronger impact at high energies, enforcing the displaced initial condition $X_n^i \simeq 0.47$. As $S$ decays away, its impact on $X_n(T)$ is even more rapidly exponentially suppressed and its constraints are less stringent than decays at very short lifetimes. The bounds from annihilation are given in the $Y_S^2 \langle \sigma v \rangle_{\pi^+ \pi^-} - \tau_S$ parameter space and displayed in Fig.~\ref{fig:Y2sigann}.

\subsection{Energy density requirements}

The resultant BBN abundances depend on the nuclear reaction rates and how efficient they are as the Universe expands. One by one, the reaction rates drop out of equilibrium, as the Universe expands and cools. 
 If the Hubble rate is increased due to a large energy density locked in a dark sector, the active reaction time would shorten, potentially spoiling the SBBN results. For our study, the most important effect is the change of the Hubble rate during the $n/p$ freeze out, which again affects $Y_p$.
 However, we can also use as a constraint a well-measured quantity in cosmology, the total energy density carried by neutrinos. 

The neutrinos decouple from thermal processes at  $T \sim 2 \MeV$. If the decaying particle is heavy and does not decay into neutrinos, it will reheat electron-photon fluid with respect to the neutrinos, decrease $T_\nu/T_\gamma$ and equivalently lower $N_{\rm eff}$. The Planck collaboration measured $N_{\rm eff}= 3.04 \pm 0.33$ at $2\sigma$, including their CMB results and external cosmological data~\cite{Ade:2015xua}, which imposes $N_{\rm eff} > 2.71$ as a lower bound. 

The energy densities and Hubble rate form a closed system of differential equations
\beq
\dot{\rho}_S + 3 H \rho_S = - \Gamma_S \rho_S, \qquad
\dot{\rho}_{rad} + 4 H \rho_{rad} = \Gamma_S \rho_S, \qquad
H^2 = \frac{8\pi G}{3}\left(\rho_{rad} + \rho_S\right), \label{eq:EnerDens}
\eeq
where we have assumed a non-relativistic $S$ and omitted the variation in relativistic degrees of freedom. Assuming step-like decoupling and changes in relativistic degrees of freedom, the $T$ evolution separates into 3 regions. For $T > T_{\nu}^{\rm decoup}$, neutrinos are in equilibrium with the electromagnetic bath and $\rho_S$ is injected equally in $e^\pm$'s, $\nu$'s and $\gamma$'s. For $T_{\nu}^{\rm decoup} > T > T_{m_e}$, the neutrinos are simply redshifted while the electron-photon bath is heated by the $S$ decays. For $T_{m_e} > T$, electrons become non-relativistic and transfer their entropy to photons, additionally heating the photon bath compared to the neutrino bath. 

If $S$ does not dominate the energy density of the Universe before its decay, we can write $\rho_S = \delta_S \rho_{rad}^{SM}$, $\rho_{rad} = \rho_{rad}^{SM}(1+\delta_{rad})$ and expand~(\ref{eq:EnerDens}) around the $\delta$ perturbations to solve the system analytically. At linear order, we find the solutions
\beq
\rho_S(t) = \frac{c_S}{t^{3/2}}e^{-\Gamma_S t}, \qquad \rho_{rad}(t) = \frac{c^i_{rad}}{t^2}\left[1 +F(t)\right],  \qquad 
F(t) = \frac{c_S}{c^i_{rad}\sqrt{\Gamma_S}}\frac{1}{\Gamma_S t} \left[ \Gamma_{3/2}(\sqrt{\Gamma_S t})- \Gamma_{5/2}(\sqrt{\Gamma_S t}) + \frac{\sqrt{\pi}}{4}\right] \label{eq:rhos}
\eeq
where $\Gamma_{3/2}$, $\Gamma_{5/2}$ are incomplete Gamma functions and the integration constants $c_S$, $c_{rad}$ are set to have $\rho_S = m_S n_S$ and $\rho_{rad} = \rho_{rad}^{SM}$ at some early time $\Gamma_S t \ll 1$. After the neutrinos decouple, the injected energy is distributed to the photon-electron bath and its energy density departs for the neutrino bath
\beq
\rho^{\rm mid}_{\gamma}(t) = \tilde{g}_{\gamma+e}\frac{c^i_{rad}}{t^2}\left[1 + F(t)\right] +  \tilde{g}_{\nu} \frac{c^i_{rad}}{t^2} \left[G(t) - G(t_\nu^{\rm decoup}) \right], \qquad 
\rho^{\rm mid}_{\nu}(t) = \tilde{g}_{\nu}\frac{c^i_{rad}}{t^2}\left[1 + F(t)-G(t) + G(t_\nu^{\rm decoup}) \right], \qquad 
\eeq
where $\tilde{g}_i \equiv g_{i}/(g_{\gamma+e}+ g_\nu)$ is the fraction of relativistic degrees of freedom of each bath, $t_\nu^{\rm decoup}$ the neutrino decoupling time and
\beq
G(t) = \frac{c_S}{2 c^i_{rad}}\sqrt{\frac{\pi}{\Gamma_S}}\erf\left(\sqrt{\Gamma_S t}\right)- \frac{c_S}{c^i_{rad}}\sqrt{t}e^{-\Gamma_S t}.
\eeq
Finally, after the electrons become non-relativistic, they effectively transfer their entropy to the photon bath. Assuming an instantaneous transition, entropy continuity implies an increase of energy density by a factor of $\delta = \left(g_e+g_\gamma\right)^{1/3}/g_\gamma^{1/3} = (11/4)^{1/3}$. Matching boundary conditions, the energy densities at late times are
\begin{align}
\rho^{\rm late}_{\gamma}(t) &= \tilde{g}_{\gamma+e}\delta\frac{c^i_{rad}}{t^2}\left[1 + \alpha F(t) + c \frac{t_e}{t}\right] +  \tilde{g}_{\nu} \alpha \frac{c^i_{rad}}{t^2} \left[G(t) - G(t_e) + \delta \left(G(t_e)-G(t_\nu^{\rm decoup})\right) \right], \\ 
\rho^{\rm late}_{\nu}(t) &= \tilde{g}_{\nu}\frac{c^i_{rad}}{t^2}\left[1 + \alpha\left\{ F(t)-G(t) + G(t_e) +\delta( G(t_\nu^{\rm decoup})- G(t_e))\right\} + c \frac{t_e}{t} \right],
\end{align}
with $\alpha = 1/(\delta \tilde{g}_{\gamma+e} +\tilde{g}_{\nu})$ and $c$ a boundary condition that is irrelevant in the $t \to \infty$ limit.

The temperature-time dependence is found via $\rho_{rad}(t) = \pi^2 g_\star T^4/30$. Since the neutrino interaction rate scales as  $\Gamma_{\nu_e}\sim T^5$, we find the neutrino decoupling time in the modified cosmology by equating $(T_\nu^{\rm decoupl})^5/H(T_\nu^{\rm decoupl}) = (T_\nu^0)^5/H_0(T_\nu^0)$, with $H$ the perturbed Hubble rate and $T_\nu^0$ the neutrino decoupling temperature in the SM. In the Maxwell-Boltzmann approximation, $T_\nu^0 = 2\MeV$, but thermal refinements in the interaction rates and phase space tend to yield a lower value $T_\nu^0 = 1.4\MeV$~\cite{Dolgov:2002wy}.

\begin{figure}
\centering
 \includegraphics[width= 0.48\columnwidth]{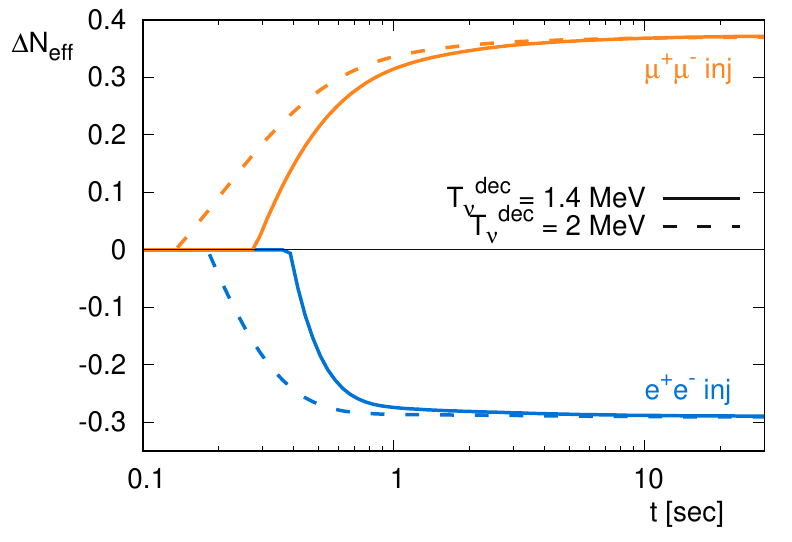} \hspace{0.5cm}
  \includegraphics[width= 0.48\columnwidth]{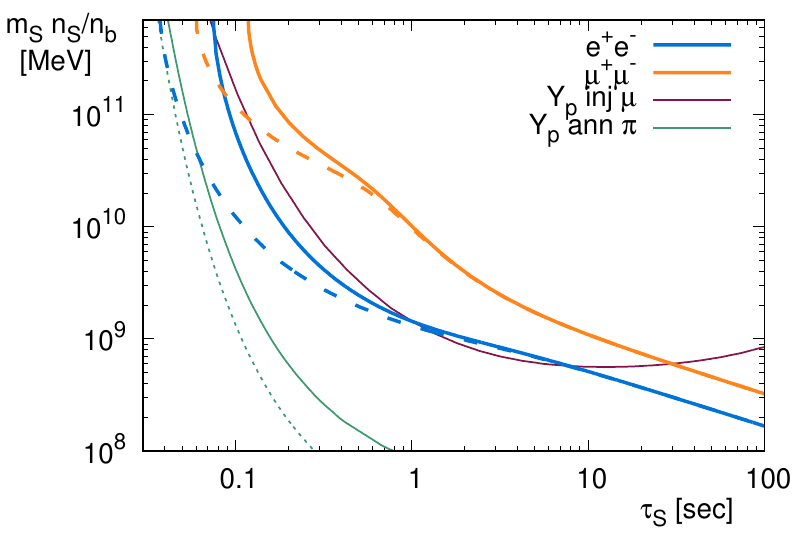}
\caption{\textit{Left}: Departure from the SM $N_{\rm eff}$ as the Universe cools down for electron injections (blue) and muon injections (orange). The extrema of the neutrino decoupling temperature ranges are shown in full lines and dashed lines as labeled in the figure. \textit{Right}: Bound of maximal stored energy decaying into electrons or muons as a function of particle lifetimes. The full line and dashed lines represent the neutrino decoupling temperatures as on the left. We also show for comparison some benchmark bounds in this parameter space from the BBN $Y_p$ results. The thin olive curves are the neutron enrichment constraint from annihilation into pions for $m_S = 140 \MeV$ (solid) and $m_S = 275 \MeV$ (dotted). The thin purple line is the $Y_p$ constraint for a $m_S = 250 \MeV$ particle decaying into muons.} 
\label{fig:Neff}
\end{figure}

Then, at $\Gamma_S t \gg 1$, we can evaluate $T_\nu / T_\gamma$ and find
\beq
N_{\rm eff} =  3 \left(\frac{T_\nu}{T_\gamma}\right)^4 \left(\frac{11}{4}\right)^{4/3} 
\simeq 3 \times\frac{\delta \tilde{g}_{\gamma + e} +\tilde{g}_\nu - \frac{c_S}{2c_{rad}^i}\sqrt{\frac{\pi}{\Gamma_S}}+ (1-\delta)G(t_e)+\delta G(t_\nu^{\rm decoup})}
{\delta \tilde{g}_{\gamma + e} +\tilde{g}_\nu  + \frac{\tilde{g}_\nu }{\delta \tilde{g}_{\gamma + e}}\left( \frac{c_S}{2c_{rad}^i}\sqrt{\frac{\pi}{\Gamma_S}}-(1-\delta ) G(t_e) - \delta G(T_\nu^{\rm decoup})\right)} 
\eeq
to constrain energy injection into electrons. We display in Fig.~\ref{fig:Neff} the departure from $N_{\rm eff} =3$ as a function of time for $\tau_S = 0.1 \sec$ and the two neutrino decoupling temperature benchmarks. The limits are also shown in units of stored energy density $m_S n_S/n_b$, where $m_S$ is in MeV. If the $S$ decay happens after the neutrino decoupling, all energy is deposited in the photon bath and the result is independent of our choice of $T_\nu^0$. If decays happen earlier, the photon and neutrinos are potentially still coupled and the energy emitted in $S$ decays only influences $N_{\rm eff}$ after decoupling. As such, the constraints has a $t\propto (1/T_\nu^0)^2$ dependence. We adopt the conservative side, $T_\nu^0 = 1.4\MeV$, as our bounds on the $m_S-\tau_S$ parameter space. Notice the constraints for $SS$ annihilations to pions are much stronger and will be dominant when the pionic annihilation channel is open, \textit{i.e.} for $m_\pi < m_S < 2 m_\pi$.

\subsubsection{Energy injection partitioned between photon and neutrino baths  (\textit{e.g. muon injection})}

The case for muon injection is somewhat interesting as its decay products, neutrinos and electrons, clearly thermalize in the two different baths, once everything is decoupled. Both $T_\gamma$ and $T_\nu$ will rise, but since the 2 neutrinos carry more energy than the electron for a muon decay, we expect a rise in $N_{\rm eff}$. More precisely, we solve a similar set of equations as~(\ref{eq:EnerDens}), except the photon bath absorbs a $\xi$ proportion of the $S$ decay energy and the neutrino bath gets the remaining $(1-\xi)$ portion. Before neutrino decoupling, the radiation bath evolves as in equation~(\ref{eq:rhos}). Each decay product carries on average the energy~\cite{Pospelov:2010cw}
\beq
\left\langle E_{e} \right \rangle = 37.0 \MeV, \qquad \qquad \left \langle E_{\nu_e} \right \rangle = 31.7 \MeV, \qquad \qquad \left \langle E_{\nu_\mu}\right\rangle = 37.0\MeV.
\eeq
After neutrino decoupling, the energetic neutrinos can still collide with the ambient electrons until $\Gamma_{{\rm coll} -e }^{\nu_e} < H$, where  $\Gamma_{{\rm coll} -e }^{\nu_e}$ is the collision rate with electrons only, the $T_\gamma$-dependent term in equation~(\ref{eq:Collnue}). Then, the energy distributed to the photon bath separates into two regimes
\beq
\xi_1 = \frac{\left \langle E_{\nu_e} \right \rangle+\frac{\Gamma_{{\rm coll} -e }^{\nu_e}}{\Gamma_{\rm coll}^{\nu_e}}\left \langle E_{\nu_e} \right \rangle + \frac{\Gamma_{{\rm coll} -e }^{\nu_\mu}}{\Gamma_{\rm coll}^{\nu_\mu}}\left \langle E_{\nu_\mu} \right \rangle}{m_\mu} \simeq 0.47, \qquad \qquad
\xi_2 = \frac{\left \langle E_{\nu_e} \right \rangle}{m_\mu} = 0.35, 
\eeq
where the muon-neutrino collision term is given by
\beq
\Gamma^{\nu_\mu}_{\rm coll} \left(E_\nu , T \right) = \frac{7\pi}{135}  \GF^2 E_{\nu} \left[ \left(5 + \left(g_{\rm L}-1\right)^2 + g_{\rm R}^2\right) T_\nu^4 + 4 \left(\left(g_{\rm L}-1\right)^2 + g_{\rm R}^2 \right) \eta(T) \;T_\gamma^4 \right].
\eeq
Following the same procedure as before, we find
\begin{align}
N_{\rm eff} &= 3 \times \frac{\delta \tilde{g}_{\gamma + e +\tilde{g}_\nu }+ \frac{c_S\left(\delta \tilde{g}_{\gamma + e}(1-\xi_2)-\xi_2\tilde{g}_\nu\right)}{2c_{rad}^2 \tilde{g}_\nu}\sqrt{\frac{\pi}{\Gamma_S}}+\frac{C}{ \tilde{g}_\nu}}
{\delta \tilde{g}_{\gamma + e} +\tilde{g}_\nu - \frac{c_S\left(\delta \tilde{g}_{\gamma + e}(1-\xi_2)-\xi_2\tilde{g}_\nu\right)}{2c_{rad}^2\delta \tilde{g}_{\gamma + e}}\sqrt{\frac{\pi}{\Gamma_S}}-\frac{C}{\delta \tilde{g}_{\gamma + e}}},\\ 
C &= \delta (\xi_2-\xi_1)G(t_{\rm coll})+ \delta (\xi_1 - \tilde{g}_{\gamma + e}) G(t_\nu^{\rm decoup})+ \xi_2 (1-\delta) G(t_e),
\end{align}
with $t_{\rm coll}$ found by solving  $\Gamma_{{\rm coll} -e }^{\nu_e} = H$. The physics is constrained by $N_{\rm eff} < 3.37$. The time-dependence of the departure from $N_{\rm eff} = 3$ is shown for $\tau_S = 0.2 \; \sec$ and the two choices of $T_\nu^0$ in Fig.~\ref{fig:Neff}. The corresponding constraints on the maximal stored energy for a given lifetime are shown on the right. For comparison with the muon-induced $Y_p$ bound, we display the curve for $m_S = 250 \MeV$ from neutron enrichment in purple. Independently from the choice of $T_\nu^0$, the bounds from $Y_p$ $\mu$-injection are most constraining for $\tau_S \gtrsim 0.2 \;\sec$ while the annihilation to $\pi^+\pi^-$ provides the dominant constraint in the entire $2m_\mu < m_S < 2 m_\pi$ range.

\subsection{Late-time energy injection }

In the example of the $S$ particles coupled through the Higgs portal, the most stringent constraints on lifetime come from the 
considerations of $n/p$ freeze out. In other models, with additional channels of annihilation that can suppress metastable abundances, 
the constraints on lifetime would not be as stringent, and would mostly come from the considerations of late 
energy injection. For completeness, we also discuss these constraints here. 
Modification of BBN by unstable particles with lifetimes in excess of 200 seconds 
has been considered in detail, both through hadronic~\cite{Reno:1987qw,Kawasaki:2004yh}, electromagnetic~\cite{Cyburt:2002uv} or combined~\cite{Kawasaki:2004qu,Jedamzik:2006xz,Pospelov:2010cw} energy cascades.

Hadronic injection after $t\gtrsim 200$ seconds is most efficient at modifying the final yields of the less abundant light nuclei D, ${}^3$He, ${}^6$Li,
and ${}^7$Li.
 After most of ${}^4$He has been synthesized, the BBN enters the regime ($T \sim 50$ keV) when neutrons are rare, $O(10^{-5})$ or so, yet their 
 abundances are critical in determining the final abundance of deuterium. At that stage, any additional neutrons brought into 
 the system through external processes such as heavy particle decays lead to the increase of the deuterium abundance. 
 (Incidentally, it also leads to the suppression of $^7$Be and consequently of $^7$Li \cite{Reno:1987qw}.)
 The increase of D production can be exacerbated by the hadro-dissociation of ${}^4$He in the  process of slowing down of injected hadrons. 
 Additional production of $^3$He through spallation can also affect the  $^3$He/D ratio \cite{Ellis:2005ii}.  
Secondary and tertiary processes may also generate  ${}^6$Li and $^9$Be \cite{Dimopoulos:1987fz,Pospelov:2010kq}.
Detailed studies of the ensuing constraints  \cite{Jedamzik:2006xz} show strong sensitivity to hadronic (mostly nucleonic) decays of metastable 
particles with lifetimes in the hundreds of seconds and longer, and initial abundances comparable or even smaller than that of baryons.
In recent years, these constraints have only got stronger, primarily due to steady observational progress in determination of primordial D/H \cite{Cooke:2013cba}. 
 
 If for some reason, hadrons and specifically nucleons are absent from the decay chains, 
 the abundances of light elements can be modified by the late injection of electromagnetic energy. 
 At early times this mechanism is inefficient, as radiation quanta with energy in excess of nuclear binding are quickly energetically degraded by 
 ambient plasma.  
 The photo-dissociation therefore sets in at late times leading to a suppression D ($t \gtrsim 10^4 $ seconds) and additional production 
 of $^3$He for $t \gtrsim 10^6$ seconds. Since typically 45\% of hadronic energy injection is dissipated electromagnetically in the hadronization cascade~\cite{Jedamzik:2006xz}, the late-time energy injection constraints on a heavy particle are dominated by the electromagnetic reactions in the BBN network.

\section{Results}

\begin{figure}
\centering
 \includegraphics[width= 0.48\columnwidth]{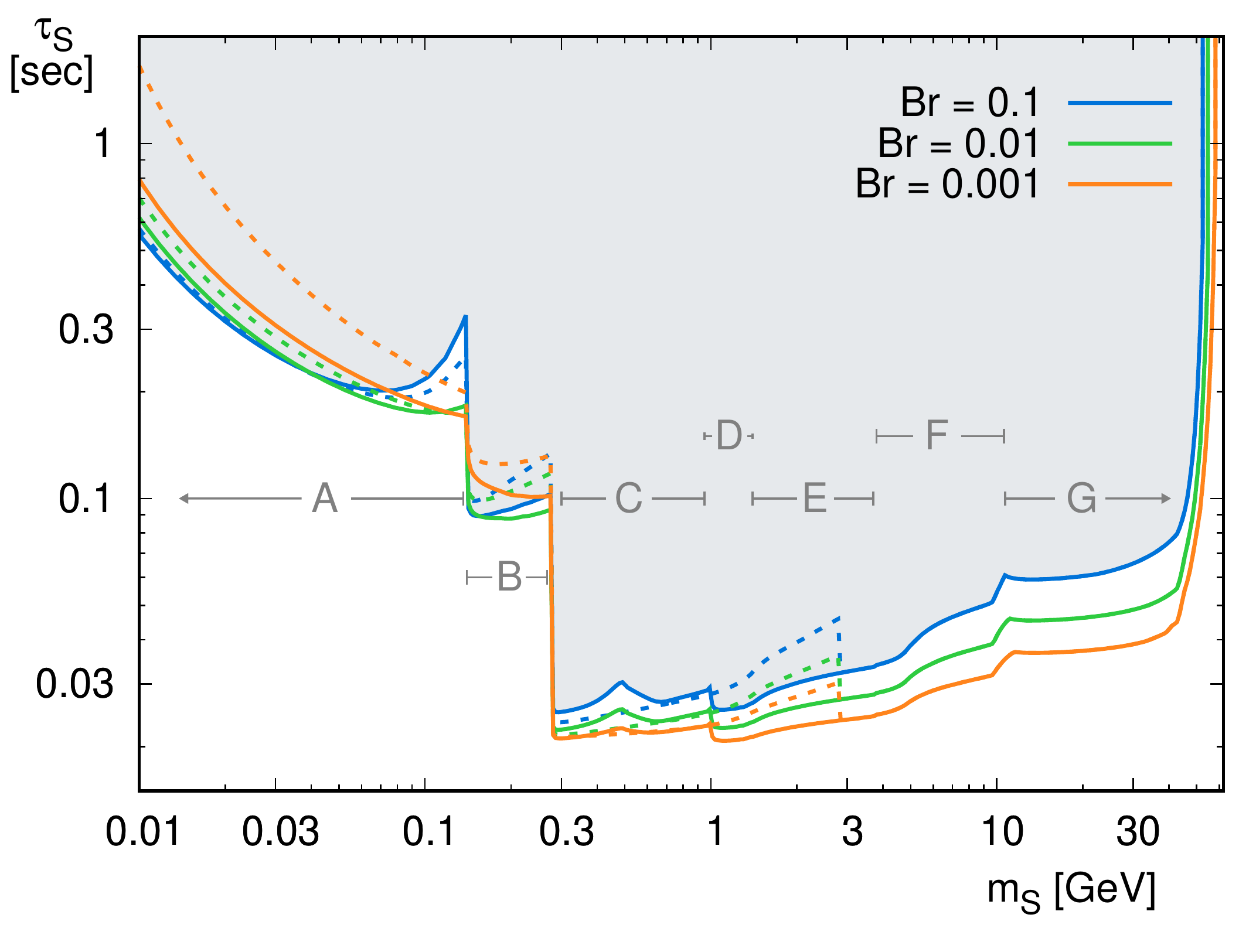} \hspace{0.5cm}
  \includegraphics[width= 0.48\columnwidth]{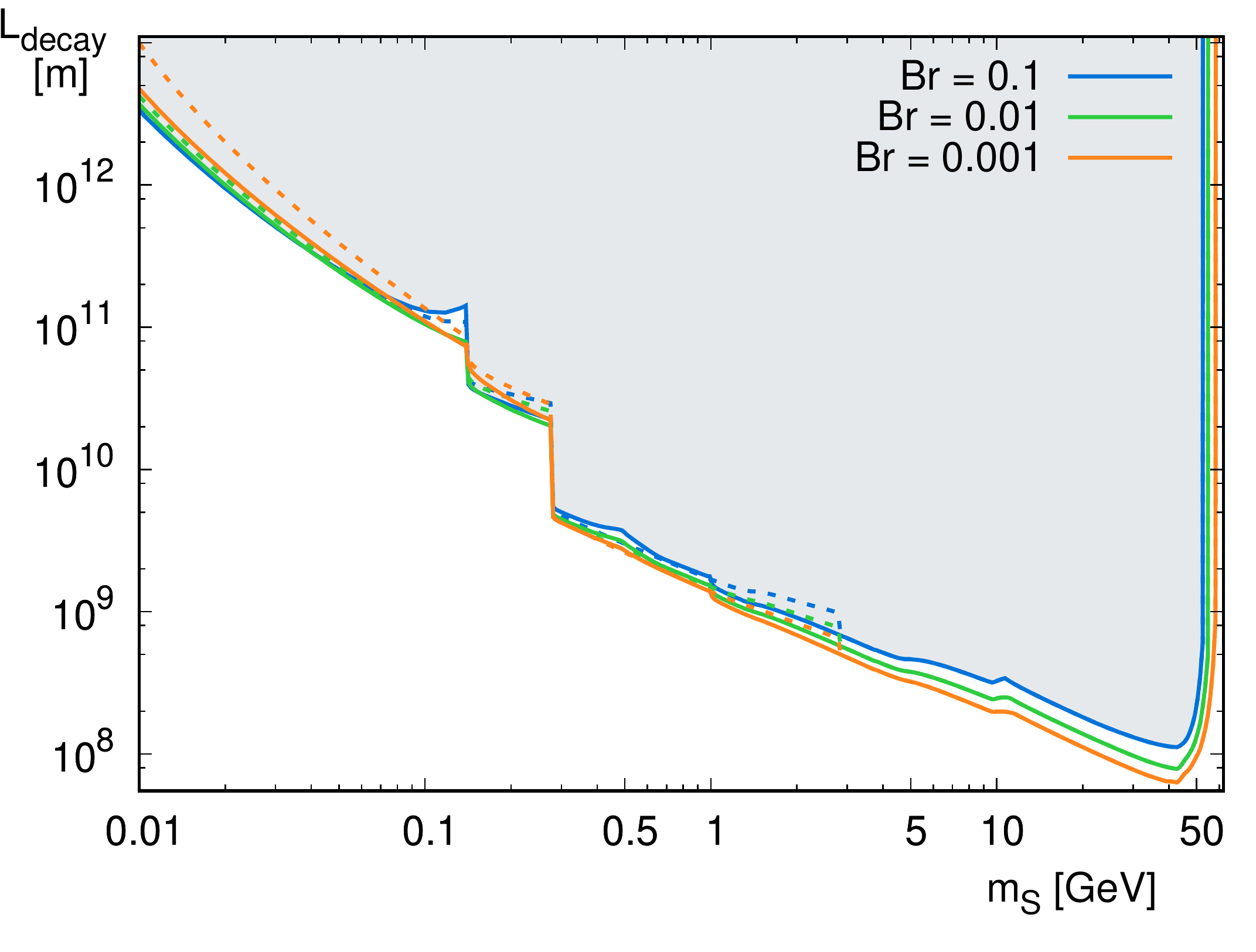}
\caption{\textit{Left}: Lifetime constraint as a function of the $S$ mass for three $h\to SS$ branching ratios. The lettered regions represent different assumptions or physics and are described in the text. The dotted lines correspond to the perturbative spectator model. \textit{Right}: Same as left, except transposed in the decay length of $S$, assuming it is boosted to  $E_S = 200\GeV$.} 
\label{fig:S_paramspace}
\end{figure}

We are now in a position to perform a scan  in parameter space of the minimal Higgs model,  constrained by the  consistency with BBN. In Fig.~\ref{fig:S_paramspace}, we display the parameter space, both in the lifetime and an effective decay length $L_{\rm dec} = c\tau_S\beta_S(E_S/m_S)$. We assume an average $E_S$ of 200 GeV, from a Higgs typically boosted at 400 GeV at the LHC.  
The resulting constraints, along with the assumptions considered in each mass range are described below.

\begin{itemize}
\item \textbf{Region A $2 m_e < m_S  < 2 m_\mu$} : The constraint comes from the decrease in $N_{\rm eff}$ with the entropy dump in the SM bath after neutrino decoupling. We take the neutrino decoupling temperature to be $T_\nu^0 = 1.4\MeV$ as a conservative limit.
\item \textbf{Region B $m_\pi < m_S  < 2 m_\pi$} : This region is dominated by the $SS$ annihilation to $\pi^+ \pi^-$. We also derived the same constraint as region A from $N_{\rm eff}$ up to $m_S =2 m_\mu$, in addition to the raised $N_{\rm eff}$ from decays into muons in the $2m_\mu < m_S < 2m_\pi$ and the $Y_p$ constraints from $S$ decaying into muons. They all yield weaker bounds, of $\tau_S > 0.3 \sec$ or longer. 
\item \textbf{Region C $2 m_\pi < m_S < 2 m_K$} : The abundance $Y_S$ weighted by the pion branching ratio constrains the region via direct charged pion decays. We assume $2/3$ go into charged pions and $1/3$ is radiated away in $\pi^0$. 
\item \textbf{Region D $2 m_K < m_S < 1.4\GeV$} : The abundance $Y_S$ weighted by the kaon branching ratio constrains the region via direct charged kaons decays. We assume $1/2$ go into charged kaons and $1/2$ into $K^0 \bar{K}^0$. Only half of the neutral kaons 
survive as $K_L$, creating  similar in numbers metastable populations of $K_L$, $K^+$ and $K^-$. 
\item \textbf{Region E $1.4\GeV < m_S < 2 m_D$} : By strangeness conservation, we assume that all $s$-quarks yield a kaon, half charged and half neutral. Since we do not have model-independent branching ratios of $S$ in this mass regime, we vary the description according to the assumptions in each decay model. For the baseline model, we assume that 100\% decays to the kaons and apply our kaon injection constraints. For the perturbative spectator model, the kaon branching ratio is given by~(\ref{eq:SpectBr}), with non-negligible contributions from decays to pions, muons and eta mesons, resulting in weaker bounds until the $c$-quark threshold. At $m_S = m_c$ the hadronic modelling dependence largely goes away.
\item \textbf{Region F $2 m_D < m_S  < 2 m_b$} : We utilize the branching fractions of $c\bar{c}$ from $e^+ e^-$ at $\sqrt{s} = 10.5\GeV$ into $D$-mesons from Ref.~\cite{Lisovyi:2015uqa} and weight each channel by its inclusive $K^\pm$ branching ratios to find a hadronization yield of 0.63 $K^+K^-$ pair per $S$ decay into $c$-quarks. Rescaled by $Br(S\to c\bar{c})$, same constraints from kaon injection apply. Above the $2m_{\Lambda_c}$ threshold, a $c\bar{c}$ typically forms a $c$-baryon with a 0.06 probability~\cite{Lisovyi:2015uqa}, which then hadronizes to $p$ or $n$. We find this constraint weaker than the kaons injection and use the $K^+K^-$ result across this entire range.
\item \textbf{Region G $m_S > 2 m_b$} : The main decay channel here are pairs of $b\bar b$ quarks. The charged pion, charged kaons and proton multiplicities in the $b\bar{b}$ decay of a $Z$ boson are measured to be $18.44\pm 0.63$, $2.63\pm 0.14$ and $1.00\pm 0.08$ respectively by the ALEPH collaboration~\cite{Barate:1997ty}. We assume the ratio holds in the hadronization of lower centre-of-mass decays into $b\bar{b}$ and scale by the mean charge multiplicity fit~\cite{Sarkisyan:2015gca}
\beq
N_{ch}(s) = -0.577 + 0.394 \ln (s/s_0) + 0.213 \ln^2(s/s_0) + 0.005 (s/s_0)^{0.55},
\eeq
where $s_0 = 1\GeV^2$. This fit agrees well in both $e^+e^-$ and $p\bar{p}$ collisions between $\sqrt{s} \sim 2\GeV-2\TeV$. This gives us an estimate for the baryon injection of the $b\bar{b}$ branching fraction of $S$. We further assume 50\% smaller injection of $n(\bar{n})$ to utilize our baryon injection constraints. The accompanying pions and kaons also independently yield comparable constraints, not shown in the figure.

\end{itemize}

\section{Discussion}

We have considered, in some detail, constraints on the lifetimes of the scalar particles, coupled to the Higgs portal via a minimal set of couplings. To stay relevant for the LHC, we have concentrated on $m_S < m_h/2$ case, that allows pair-production of $S$ 
states in the decay of Higgs bosons. The same coupling is responsible for the cosmological depletion of 
$S$ particles, leading to their metastable abundance in the early Universe. 

We find that throughout almost the whole mass range considered in this work, $2m_\mu < m_S < m_h/2$, 
the constraints on the lifetime of $S$ particles are stronger than 
$0.1$ seconds. Moreover, the results have a relatively mild dependence on the $Br(h\to SS)$. The reason for that is as follows:
the experimental limits on $Br(h\to SS)$ are already strong enough to limit the annihilation rate of $SS$ pairs to the SM states to be much less 
than one picobarn, and consequently the metastable 
abundance of $S$ particles per nucleon is quite high, $Y_p \gg 1$. This leads to a massive injection
of nucleons and mesons at early times, which raises the $n/p$ ratio, and creates larger yields of $^4$He compared to SBBN. Contributions of 
very light $S$ particles to the Hubble rate during the $n/p$ freeze out also raises $Y_p$.  The limits on $\tau_S$ 
are robust, and have rather mild dependence on the uncertainties in our treatment. This is because the initial large metastable 
$Y_S$ abundance needs to be depleted prior to the $n/p$ freeze out time $t_{n/p}$, leading to the requirement $\tau_S \ll t_{n/p}$. 
Consequently $O(1)$ variations in the yields of mesons and nucleons in the final states can be compensated by small variations in $\tau_S$, parametrically on the order $\log^{-1}(t_{n/p}/\tau_S)$,  to produce the same influence on BBN. For the same reasons, our limits are also very insensitive to the exact  observational constraint on $\Delta Y_p$, and we take a rather conservative limit of 0.01 (allowing $\pm0.01$ deviations frrom the observed/calculated mean). From the point of the LHC physics, the most promising is a scenario with a mass 
$m_S$ not far below $m_h/2$. In that case, the effective decay length has to be on the order or smaller than $\sim 10^8$ meters, Fig.~\ref{fig:S_paramspace}, providing a 
$10^{-6}$ {\em minimum} probability for a decay within a 100 m length purposely built detector. 
Given that the high-luminosity LHC would produce copious numbers of the Higgs bosons, there is a chance to cover the entire 
lifetime range for masses within $10-$to$-50$ GeV range.

It is easy to see that the above considerations can be generalized to other models of the Higgs-portal-coupled particles. For example, 
consider a fermion $\chi$, coupled to the Higgs via $H^\dagger H (\bar\chi \chi)$ or $H^\dagger H(\bar \chi i \gamma_5 \chi)$ dimension-five operators, and having a small decay term such as {\em e.g.} neutrino portal $LH \chi$. The main 
analysis of our work can be recast for that model, especially in the part that connects Higgs 
decays with a metastable abundance of $\chi$. Evidently, for $Br(h\to \chi\bar\chi ) \sim Br(h\to SS) $ input, one will end up with $Y_\chi \sim Y_S$.  The only change will be in the yields of mesons and baryons in the decays of $\chi$ 
compared to $S$. However, it is well known that already for $m_\chi $ above 250 MeV, the yields of pions and kaons is substantial 
\cite{Alekhin:2015byh}, giving confidence that for the most parts same constraints we have derived for $\tau_S$ will translate to similar limits 
on $\tau_\chi$. 

The analysis performed in this paper can be easily generalized to other models of metastable particles, with different types of interactions, via $Z$, $Z'$ etc. In the limit when $Z'$ is outside of the LHC reach, one could have a set of effective operators connecting $\chi$ with 
the SM fields, such as $\frac{1}{\Lambda^2}\bar \chi \gamma_\mu \chi \bar q \gamma_\mu q$, where $\Lambda$ is some energy scale. 
The $\chi$  pair-production cross section in this case will scale as $\sigma_{q\bar q \to \chi \bar\chi } \propto E_q^2 \Lambda^{-4}$, where 
$E_q$ is a typical (anti-)quark energy, while cosmological annihilation cross section has $\sigma_{\chi \bar\chi \to q\bar q  } v \propto 
m_\chi^2 \Lambda^{-4}$ scaling. Therefore, the LHC-relevant cross section can be enhanced relative to the annihilation 
rate by a parametrically large ratio, $ E_q^2/m_\chi^2  $  if $m_\chi$ is parametrically smaller than the TeV scale. Therefore, 
one can easily have a range of parameters with a relatively large $\chi\bar\chi$ pair-production cross section, while having very small annihilation 
rates, rendering $Y_\chi \gg 1$, and resulting again in strong BBN constraints on lifetimes, $\tau_\chi < 0.1$ seconds. 
Therefore, we conclude that some simple $Z'$ mediated models of metastable particles can also be strongly restricted by cosmology, making 
them a perfect candidate for the searches of metastable particles at the LHC.

It is also instructive to consider models where constraints on the lifetime of metastable particles are {\em much weaker}. Clearly, one needs 
an effective new mechanism for the self-annihilation in the early Universe, as the Higgs channel is too inefficient. Staying within 
the Higgs portal models, consider the following potential with two real scalars,
\begin{equation}
V(H,S_1,S_2)  =  H^\dagger H (\lambda_1 S_1^2 + \lambda_2 S_2^2+ A_1S_1 + A_2S_2) +\lambda_{12}S_1^2S_2^2 + V(S_1) + V(S_2) + V(H^\dagger H),
\end{equation}
with the following hierarchy of couplings: 
\begin{equation}
\lambda_1 \gg \lambda_2;~ A_1 \ll A_2; ~ \lambda_{12} \sim O(1);~ m_{S_1} > m_{S_2}. 
\end{equation}
These choices will lead to a long-lived $S_1$, somewhat shorter-lived $S_2$, a predominant decay of 
the Higgs boson to pairs of $S_1$, and cosmological 
depletion of $S_1$ via $S_1S_1 \to S_2 S_2$ annihilation with potentially a large cross section due to a sizeable $\lambda_{12}$ 
coupling.  Most importantly, in this model the Higgs decay to pairs of $S_1$ does not result in a prediction of $Y_{S1}$ abundance, 
which can be quite small even for small values of $Br(H\to S_1S_1)$. 
If $Y_{S1} \ll 1$, there would not be enough decay mesons and nucleons to affect early $n/p$ freeze out, and  constraints on $\tau_{S1}$ will be coming only from the considerations of late decays 
with hadronic or electromagnetic energy injection. Instead of $\tau_{S}<0.1$ sec, one expects to have sensitivity to $\tau_{S1} \sim 10^3$ seconds, or even worse, beyond $10^4$ seconds, if decays of $S_1$ are mostly leptonic. This example is not unique, and 
there are other models where constraints on lifetimes and decay lengths are relatively lax, provided that there are extra channels that ensure 
efficient cosmological annihilation of metastable particles.

\section*{Acknowledgements}
We thank D. Curtin, M. McCullough, P. Meade, M. Papucci and J. Shelton for soliciting this study, as well as  D. Curtin and B. Shuve for
very  helpful discussions.  Research
at Perimeter Institute is supported by the Government
of Canada through Industry Canada and by the
Province of Ontario through the Ministry of Economic
Development \& Innovation.

\end{document}